\documentclass{article}
\input jetpl_23.sty
\lat
\sod{Spin - glass transition in Kondo lattice with quenched disorder}
{Kiselev M.N., Oppermann R.}

\begin{document}

\author{M.N.Kiselev$^{+*}$ and R.Oppermann$^{+\S}$}
\address{$^+$Instit\"ut f\"ur Theoretische Physik, Universit\"at W\"urzburg,
D-97074 W\"urzburg, Germany\\
$^*$ Russian Research Center "Kurchatov Institute", 123 182 Moscow, Russia\\
$^\S$University of Oxford, Department of Physics, 1 Kable Road, OX1 3NP Oxford, UK}
\date{January 21, 2000}
\title{Spin - glass transition in Kondo lattice with quenched disorder}
\maketitle

\begin{abstract}
We use the Popov-Fedotov representation of spin operators to construct
an effective action for a Kondo lattice model with quenched disorder
at finite temperatures. We study the competition between the Kondo
effect and frozen spin order in Ising-like spin glass. We
present the derivation of new mean-field equations for the spin-glass order
parameter and analyze the effects of screening of localized spins
by conduction electrons on the spin-glass phase transition.
\end{abstract}


One of the most interesting questions of physics of heavy-fermion
(HF) compounds is the competition between Kondo screening of
localized spins by conduction electrons (CE) and ordering of these
spins due to Ruderman-Kittel-Kasuya-Yosida (RKKY) interaction
(see, e.g. \cite{GS}). The screening is attributed to the Kondo effect
- the resonance scattering of an electron on a magnetic atom with
simultaneous change of the spin projection. In dilute alloys such
a scattering results in the sharp resonance at Fermi level with
characteristic energy width $\epsilon \sim T_K \sim
\epsilon_F\exp(-\alpha^{-1})$, where $T_K$ is Kondo temperature,
 $J$ is a coupling constant, $\rho$ is a
density of states of CE on the Fermi level and $\alpha=\rho J$.
As it was recently discussed (see, e.g. \cite{R1}-\cite{R2}), such
a competition can be responsible for the Non-Fermi-liquid (NFL)
behaviour observed in some heavy fermion compounds. Most of such a
materials share two characteristics: proximity to magnetic region
of appropriate phase diagram (usually temperature vs. pressure or
chemical composition), and disorder due to chemical substitution.
In many respects the concentrated Kondo systems, namely the
lattice of magnetic atoms interacting with CE
"bath" (Kondo lattice (KL)) show striking similarities with dilute
Kondo systems. The Kondo temperature in these systems is a
characteristic crossover temperature at which spins transform
their local properties to some itinerant Fermi-liquid behaviour
determining low temperature regime of HF compounds. NFL behaviour
in HF system is mainly attributed then to reducing the Kondo
temperature possibly even suppressing it to zero. In turn, magnetic
or spin glass (SG) transition can also be suppressed due to
interplay between Kondo scattering and spin-spin interaction.
Thus, such an interplay can result in  quantum phase transition
(QPT) \cite{R1} when both Kondo and magnetic temperatures are
equal to zero at some finite doping. The role of chemical
substitution in this case is to "tune" the Fermi level of metallic
system providing sharp Kondo resonance.

The problem of competition between RKKY and Kondo interaction in
clean system was studied for the first time by Doniach \cite{Don}
in the "Kondo necklace" model. The transition typically takes place between
a paramagnetic metal and magnetic (usually AFM) metal. In this case there are
two possibilities: the compound will have long range magnetic
order when the RKKY interaction is sufficiently large compared
with the Kondo interaction, or the compound will be paramagnetic
due to the quenching of magnetic moments of the rare earth atoms
and the ground state has the features of Kondo-singlet state.
Nevertheless, in the region  $T^M_{RKKY} \sim T_K$ the competition between
magnetic and Kondo interaction results in dramatically change of
"naive" Doniach diagram (see \cite{KKM}). Namely, both Kondo and
magnetic temperatures are strongly suppressed and spin-liquid
state (e.g of Resonance Valence Bond type \cite{CA}) occurs.

The goal of this letter is to present some results concerning the
competition between Kondo effect and Ising-like  SG transition
which is in many aspects similar to the magnetic instability. We
study mechanisms of suppressing the SG transition and effects of
screening in disordered environment. In this paper we consider the 
high temperature regime of KL model. We leave aside the issue of the
ground state properties and especially the question whether NFL
behaviour is a generic feature of vicinity to QPT for future
publication.

The Hamiltonian of KL model with additional quenched randomness of
exchange interaction between localized spins is given by
\begin{equation}
H_{KL}=\sum_{k\sigma}\varepsilon_k c^\dagger_{k\sigma}c_{k\sigma}
+J\sum_i\left(\vec{s}_i\vec{S}_i +\frac{1}{4}n_iN_i\right)
-\sum_{ij}I_{ij}\left(S^z_iS^z_j+\lambda
S_i^+S_j^- \right) \label{ham} \end{equation}
The system
under consideration is a periodic lattice of magnetic atoms
modeled by $f$ - orbitals interacting with metallic background
spin density operator $\vec{s}_i=\frac{1}{2}c^\dagger_{i\alpha}
\vec{\sigma}_{\alpha\alpha'}c_{i\alpha'}$. The first term
in the Hamiltonian (\ref{ham}) describes kinetic energy of
CE, the second stands for Kondo coupling
($J>0$). We denote
$n_i=\sum_\sigma c^\dagger_{i,\sigma}c_{i,\sigma}$ as the CE
density operator.
The identity $N_i=1$ describes the half-filled f-electron
shell. Quenched independent random variables $I_{ij}$ with
distribution $P(I_{ij}) \sim \exp (-I^2_{ij}N/(2I^2))$ stand for
direct spin-spin interaction \cite{Op1}. We assume that this
random interaction is of RKKY origin
\footnote{It has been pointed out in [8], that the
presence of nonmagnetic impurities makes RKKY interaction
a random interaction even in the case of regular arrangement of magnetic 
moments.}, namely, for $d$-dimensional
system $I \sim \alpha^2\epsilon_F l^{-d}$, $l$ is a lattice
constant in magnetic sublattice. The magnetic effects can also be
included in our approach by introducing nonzero standard deviation
$\Delta I = \bar I_{RKKY}$ into the distribution $P(I_{ij})$,
which, in turn, can result in the additional competition between
SG and AFM (or, rarely FM) states. For simplicity we neglect these
effects in present letter concentrating on the interplay between
Kondo interaction and effects of bond disorder. Since indirect
RKKY interaction through CE is mostly determined
by "fast" electrons with characteristic energies $\epsilon \sim
\epsilon_F \gg T_K$ we neglect also the Kondo renormalizations of
RKKY exchange.

As it well-known for a long time, the spin $S=1/2$ matrices can be
exactly replaced by bilinear combination of Fermi operators
$S^z_i=\frac{1}{2}(f^\dagger_{i\uparrow}f_{i\uparrow}-
f^\dagger_{i\downarrow}f_{i\downarrow})$,
$S^+_i=f^\dagger_{i\uparrow}f_{i\downarrow}$,
$S^-_i=f^\dagger_{i\downarrow}f_{i\uparrow}.$ Nevertheless, most of
fermionic representations of spin are not free of constraint
problem. For this reason, the dimensionality of space in which
these operators act is always greater than the dimensionality of
the spin matrices. Elimination of unphysical states is a serious
problem which makes the diagrammatic techniques quite complicated.
Moreover, in most cases, the  analytical continuation of Feynman
diagrams becomes extremely uneasy. To avoid the main part of
difficulties related to constraint, the new representation for
spin operators was proposed in well-forgotten paper of Popov and
Fedotov \cite{Popov}. In this representation the partition
function of the problem containing spin operators ($H_S$) then can
be easily expressed in terms of new fermions with imaginary
chemical potential ($H^f_S$):
$
Z_S={\tt Tr}\;e^{-\beta H_{S}}=
i^N {\tt Tr}\;e^{-\beta(H^f_{S}+i\pi N_f/(2\beta))}$,
$N_f=\sum_{i\sigma} f^\dagger_{i\sigma}f_{i\sigma}$ and $\beta=1/T$.
As a result, there is no constraint, the unphysical states are eliminated  and
standard Matsubara-Abrikosov-Gor'kov diagrammatic technique is obtained
\cite{Popov},\cite{Op2},\cite{BK}.

We sketch our derivation of the effective action and of resulting mean field (MF)
equations for KL model in order to explicit the approximations made and the physics
underlying these approximations. To
construct the path integral representation for the partition
function, the new Grassmann variables $c^\dagger_{i\sigma} \to
\bar{\Psi}_{i\sigma},\;\; c_{i\sigma} \to \Psi_{i\sigma}$ for
CE  with chemical potential $\mu$ and
$f^\dagger_{i\alpha} \to \bar{a}_{i\alpha},\;\; f_{i\alpha} \to
a_{i\alpha}$ for Popov-Fedotov (PF) spin operators ($S=1/2$) are
introduced. The Euclidean action for the KL model is given by
\begin{equation}
{\cal A} =\int_0^\beta d\tau \left(\sum_{i\alpha}\left[
\bar{\Psi}_{i\alpha}(\tau)(\partial_\tau
+\mu)\Psi_{i\alpha}(\tau)+
\bar{a}_{i\alpha}(\tau)(\partial_\tau-i\pi T/2)
a_{i\alpha}(\tau)\right] - H_{int}(\tau)\right).
\label{eff1} \end{equation} where the generalized Grassmann fields satisfy
the following boundary conditions:
$
\Psi_{i\alpha}(\beta)=-\Psi_{i\alpha}(0)\;$,
$\bar{\Psi}_{i\alpha}(\beta)=-\bar{\Psi}_{i\alpha}(0)$,
$\;a_{i\alpha}(\beta)=i a_{i\alpha}(0)$,
$\;\bar{a}_{i\alpha}(\beta)=-i\bar{a}_{i\alpha}(0)$.

In this paper we consider $\lambda=0$ which corresponds to
Sherrington- Kirkpatrik (SK) \cite{SK} spin-glass model. Such an
anisotropy of RKKY interaction can be associated e.g. with lattice
geometry. In
the case of Ising-like model the dynamical fluctuations in spin
subsystem appear only due to interaction with conduction electrons
and in high temperature regime $T \sim T_{SG}$ can be neglected.
To study the influence of Kondo-scattering on SG transition
temperature $T_{SG}$ we use standard replica trick
$\Psi_i(\tau)\;\to\; \upsilon^a_i(\tau),\;
a_i(\tau)\;\to\;\varphi^a_i(\tau),\;\;a=1,...n.$ Then, the free
energy of the model can be calculated (see, e.g. \cite{Binder}) by
taking the formal limit $n\to 0$ in
\begin{equation}
<Z^n>_{av}=\prod\int dI_{ij} P(I_{ij})\prod
D[\varphi^a_{i,\sigma}\upsilon^a_{i,\sigma}]
\exp\left({\cal A}_0[\upsilon^a,\varphi^a]-
\int_0^\beta d\tau H_{int}(\tau)\right)
\label{part2}
\end{equation}
where ${\cal A}_0$ is corresponding to noninteracting fermions.

As we already mentioned, for considering
the competition between Kondo scattering and trend of
disorder we assume that the magnetic temperature
$T^M_{RKKY}\ll T^*$, where $T^*$ stands for
characteristic temperature corresponding to the Kondo temperature
in a lattice. This assumption allows one to decouple the Kondo
interaction term
$H^K_i=-\frac{J}{2} \displaystyle 
\bar\upsilon^a_{i,\sigma}\varphi^a_{i,\sigma}
\bar\varphi^a_{i,\sigma'} \upsilon^a_{i,\sigma'}$ in each site
by the replica-dependent Hubbard-Stratonovich field $\psi^a_i$ \cite{RN}.
Performing the average over random potential in Eq.(\ref{part2}) results in
\begin{multline}
<Z^n>_{av}=\prod\int D[\upsilon^a,\varphi^a,\psi^a]
\exp\left({\cal A}_0+\frac{I^2}{4N}{\tt Tr}[X^2] +\right.\\
\left.+
\int_0^\beta d\tau\sum_{i,a,\sigma}
\left\{\psi^a_i\bar\upsilon^a_{i\sigma}\varphi^a_{i\sigma}+
\psi^{a*}_i\bar\varphi^a_{i\sigma}\upsilon^a_{i\sigma}
-\frac{2}{J}|\psi^a_i|^2\right\}\right)
\label{part3}
\end{multline}
with $X^{ab}(\tau,\tau')=\sum_i\sum_{\sigma,\sigma'}
\bar\varphi^a_{i,\sigma}(\tau)\sigma\varphi^a_{i,\sigma}(\tau)
\bar\varphi^b_{i,\sigma'}(\tau')\sigma'\varphi^b_{i,\sigma'}(\tau')$.
The next step is to perform the Gaussian integration over the
replica-dependent Grassmann field $\upsilon^a$ describing
CE and to decouple the eight-fermion term ${\tt Tr}[X^2]$
with the help of $Q$ matrices (see details in \cite{Op2}). As a result, the
partition function is given by:
\begin{multline}
<Z^n>_{av}=\int D[Q]\exp\left(-\frac{1}{4}(\beta I)^2N{\tt Tr}[Q^2]+\right.\\
\left.+
\sum_i\ln\left\{\prod\int D[\varphi^a,\psi^a]\exp\left[\sum_a\sum_{\{\omega\}}
\bar\varphi^a_{i\sigma}{\cal G}_a^{-1}\varphi^a_{i\sigma}+\frac{1}{2}(\beta I)^2{\tt Tr}[QX]\right]
\right\}\right)\end{multline}
where ${\cal G}^{-1}$ is inverse Green function for PF fermions depending on
Matsubara frequences $\omega_n=2\pi T(n+1/4)$ (see details in \cite{Popov})
\begin{equation}
{\cal G}_a^{-1}=
i\omega_{n_1}\delta_{\omega_{n_1},\omega_{n_2}} -T\sum_\epsilon
\psi^{a*}_i(\epsilon_l+\omega_{n_1})G_0(-i\nabla_i,\epsilon_l)
\psi^a_i(\epsilon_l+\omega_{n_2})
\label{GF} \end{equation} and $G_0(-i\nabla,\epsilon_l)=(i\epsilon_l -
\varepsilon(-i\nabla)+\mu)^{-1}$ stands for CE Greens
function, $\epsilon_l=2\pi T(l+1/2)$.

 We are still left with a term of fourth order
residing in ${\tt Tr}[QX]$ and can not evaluate the Grassmann
integral directly. Consequently, the second decoupling is needed.
To perform it, we stress that we do not intend to deal with
dynamical behaviour here confining ourselves by high temperature
regime in the vicinity of SG transition such that the lowest
Matsubara frequency is sufficient. Assuming this and
recalling that the spatial fluctuations are suppressed by the
choice of infinite range interaction \cite{SK}, one can consider
$Q$ as a constant saddle point matrix under condition $Q=Q^T$. The
elements of this matrix will later be determined self-consistently
from the saddle point condition. Assuming that the elements of $Q$
are $Q_{SP}^{aa}=\tilde q$ and $Q_{SP}^{a\ne b}=q$ one can
decouple the ${\tt Tr}[QX]$ term by introducing
replica-independent $z$ and replica-dependent $y^a$ fields and map
KL problem with disorder onto effective one-site interacting spin
system coupled with external local replica-dependent magnetic
field:
\begin{multline}
<Z^n>_{av}=\exp\left(-\frac{1}{4}(\beta I)^2N(n\tilde q^2+n(n-1) q^2)+\right.\\
\left.+\sum_i\ln\left[\prod\int D[\varphi^a,\psi^a]\int_z^G\int_{y^{a}}^G
\exp({\cal A}[\varphi^a,\psi^a,y^a,z])\right]\right)
\label{part4}
\end{multline}
where $\int_z^G f(z)$ denotes
$\int_{-\infty}^{\infty}dz/\sqrt{2\pi} \exp(-z^2/2) f(z)$,
\begin{equation}
{\cal A}[\varphi^a,\psi^a,y^a,z]=\sum_{a,\sigma}
\bar\varphi^a_\sigma\left[{\cal G}_a^{-1}-\sigma H(y^a,z)\right]
\varphi^a_\sigma-\frac{2}{J}\sum_\omega|\psi^a(\omega)|^2
\label{eff2} \end{equation} and $H(y^a,z)=I\sqrt{q}z+I\sqrt{\tilde q - q}y^a$
is effective local magnetic field. Note, that the variable
$q=<S_i^aS_i^b>$ corresponds to Edwards-Anderson SG order
parameter when the limit $n\to 0$ is taken. Nevertheless, the
diagonal element $\tilde q$ can be set neither zero nor one, in
contrast to the classical Ising glass theory because of dynamical
effects due to the interaction with CE "bath". To take
into account this interaction we include replica dependent
magnetic field into bare Green's function ${\cal
G}^a_{0\sigma}=(i\omega_n - \sigma H(y^a,z) )^{-1}$ and perform the
integration over PF Grassmann variables with the help of the
expression
\begin{equation}
{\tt Tr}\ln\left({\cal G}_a^{-1}-\sigma H\right)=
\ln\left(2\cosh(\beta H)\right)+ {\tt
Tr}\sum_{m=1}^\infty\frac{(-1)^{m+1}}{m} \left({\cal
G}^a_{0\sigma}(H)\Sigma(\psi^a)\right)^m \label{spur} \end{equation} where
$\Sigma(\psi^a)=-T\sum_\epsilon
\psi^{a*}_i(\epsilon+\omega_{n_1})G_0(-i\nabla_i,\epsilon)
\psi^a_i(\epsilon+\omega_{n_2})$
depends on the variable $\psi$ "responsible" for Kondo
interaction.  Calculating the first term in expansion (\ref{spur})
one gets the following expression for the effective "bosonic"
action in the one loop approximation
\begin{equation}
{\cal A}[\psi^a,H]=\ln\left(2\cosh(\beta H(y^a,z))\right)
-\frac{2}{J}\sum_n
\left[1-J\Pi(i\Omega_n,H(y^a,z))\right]|\psi^a|^2-O(|\psi^a|^4)
\label{eff3}
\end{equation}
The polarization operator $\Pi$ in the limit $T,H \ll \epsilon_F$
is given by:
\begin{multline}
\Pi(i\Omega_n,H)=-\beta^{-1}\sum_{n,\vec{k},\sigma}
G_0(\vec{k},i\epsilon_n+i\Omega_n){\cal G}_{0\sigma}(i\epsilon_n,H)
\stackrel{\Omega_n=0}{\longrightarrow}\\
\stackrel{\Omega_n=0}{\longrightarrow}
\rho(0)\left[\ln\left(\frac{\epsilon_F}{\sqrt{H^2+\pi^2\beta^{-2}/4}}\right)+
\frac{\pi}{2\cosh(\beta H)}
+O\left(\frac{H^2}{\epsilon^2_F}\right)\right]
\label{Pi}
\end{multline}
When $H=0$, the coefficient in front of $|\psi^a|^2$ in (\ref{eff3}) changes its sign at
$T^*\sim \epsilon_F\exp(-\alpha^{-1})$. This is a manifestation of
single-impurity Kondo effect (see, e.g. \cite{RN},\cite{WT}).

One can now perform the Gaussian integration over $\psi^a$
fields in (\ref{part4}) by stationary phase method
$\int D[\psi^a]\exp(\delta{\cal A}[\psi^a])=\exp\left(-{\tt Tr}
\ln\left[1-J\Pi(i\Omega_n,H(y^a,z))\right]\right)$.
After the last step, namely integration over replica dependent
field $y^a$ the limit
$n\to 0$ can be taken. The
free energy per site $f=\beta^{-1}\lim_{n\to 0}(1-<Z^n>_{av})/(nN)$ is given by
\begin{equation}
\beta f(\tilde q, q)=\frac{1}{4}(\beta I)^2\left(\tilde q^2-
q^2\right)- \int_z^G\ln\left( \int_y^G\frac{2\cosh(\beta
H(y,z))}{1-J\Pi(0,H(y,z))}\right). \label{MF} \end{equation} New
equations for $q$, $\tilde q$  are determined by conditions
$\partial f(\tilde q, q)/\partial\tilde q=0$, $\partial f(\tilde
q, q)/\partial q=0$:
\begin{equation}
\frac{1}{2}(\beta I)^2\tilde q= \int_z^G\frac{\partial\ln{\cal
F}}{\partial \tilde q},\;\;\;\; \frac{1}{2}(\beta I)^2 q=
-\int_z^G\frac{\partial\ln{\cal F}}{\partial q},\;\;\;\; {\cal
F}=\int_y^G\frac{2\cosh(\beta H(y,z))}{1-J\Pi(0,H(y,z))}
\label{MF1} \end{equation} The equations (\ref{MF} - \ref{MF1}) contain the key
result of the paper. They represent the solution of KL problem
with quenched disorder on a replica symmetrical level. To
demonstrate some interesting physical effects, described by these
equations let us consider the case $T\sim T_{SG}$$\geq$$T^*$ (Kondo high
temperature limit). Since $H(y^a,z)$ is dynamical variable, we break the
parametrical region of $H$ to several pieces. First, when $H$$\gg$
$T$,$T^*$, the logarithm in (\ref{Pi}) is cut by $H$ and there are no
temperature dependent Kondo corrections to the MF equations. This
corresponds to the limit $T^*$ $\ll$ $I$ providing frozen spins and
preventing them from resonance scattering
\footnote{We also note, that when $T^*$ $\gg$ $I$ 
the SG transition does not happen.}.
 Nevertheless, when $T^*
\sim I$, the region $H$ $\leq$ $T$ becomes very important. We calculate
${\cal F}$ expanding the r.h.s of (\ref{MF})  up to $(H/T)^2$
\begin{equation}
\ln( C{\cal F}_{z,\tilde q, q})=\displaystyle
-\frac{1}{2}\ln\left(1+\gamma u^2r^2\right)+
\frac{u^2}{2} \frac{r^2 - q\gamma z^2} {1+\gamma
u^2r^2} +\ln\left[ \cosh\left(\frac{u z
\sqrt{q}}{1+\gamma u^2r^2} \right)
\right]\label{F} \end{equation} We use the  following
short-hand notations: $u=\beta I$, $\gamma=2c/\ln(T/T^*)$,
$r^2=\tilde q - q$, $C=2c\alpha/\gamma$ with
$c=(\pi/4+2/\pi^2) \sim 1$. We note again that when $J=0$, which
corresponds to the absence of Kondo interaction, ${\cal
F}(z,\tilde q, q)= \exp\left(\frac{1}{2}(\beta I)^2(\tilde q -
q)\right)\cosh(\beta Iz\sqrt{q})$ and standard SK equation
\cite{SK} takes place, providing, e.g. an exact identity $\tilde q
= 1$.

In the vicinity of the phase transition point Eq.(13) reads:
\begin{multline}
\tilde q = 1-\frac{2c}{\ln(T/T^*)} +
O\left(\frac{1}{\ln^2(T/T^*)}\right),\\
q=\int_z^G\tanh^2\left(\frac{\beta I z \sqrt{q}} {1+2c(\beta
I)^2(\tilde q - q)/\ln(T/T^*)}
\right)+O\left(\frac{q}{\ln^2(T/T^*)}\right) \label{MF2}
\end{multline}
These
equations describe a second-order SG transition in Ising-like SK
\footnote{When an Ising system described by (1) with nearest neighbor 
interaction is treated with MF theory, equations identical to (13) are
obtained with $\sqrt{Z}I$ replacing $I$, 
where $Z$ is average number of neighbors.}
system coupled with CE "bath" in the presence of
Kondo scattering. Taking the limit $q\to 0$ we estimate the
temperature of SG transition $(T_{SG}/I)^2=1-4c/\ln(T_{SG}/T^*)-
... <1$. Thus, the Kondo scattering resonance results in
depressing of SG transition temperature due to the screening
effects in the same way as magnetic moments and one-site
susceptibility are screened in single-impurity Kondo problem
\cite{WT}. This screening shows up at large time scale
$t$ $\geq$ $1/ T^*$ and affects both diagonal and nondiagonal 
elements of $Q$ matrix. Moreover, $\tilde q$ becomes partially screened
well  above the SG transition point.
 Recalling that $H$ $\sim$ $Iy\sqrt{\tilde q}$ one can see
that our assumption $H/T$ $\leq$ $1$ is consistent with Eq.(\ref{MF2})
even if $T$ $\sim$ $T_{SG}$.
It is necessary to note, that a growing SG order parameter in
Eq.(\ref{eff3}-\ref{Pi}) suppresses Kondo effect as well as providing
a broader validity domain for equations (\ref{MF2}). We leave the
self-consistent analysis of Eq.(\ref{MF},\ref{MF2}) for a future
detailed publication.

In conclusion, we have considered the Kondo high temperature limit
(in a sence of $T>T^*$) of a KL model with
quenched disorder. We derived new MF equations for SG transition in the
presence of strong Kondo scattering and have shown that the partial screening
of both diagonal and nondiagonal elements of $Q$ matrix takes place.
As a result, the temperature of SG transition is strongly suppressed
when Ising and Kondo interactions are of the same order of magnitude.

We thank F.Bouis, B.Coqblin, K.Kikoin, and P.Pfeuty
for useful discussions. This work is
supported by the SFB410 (II-VI semiconductors).
One of us (MNK) is grateful to Alexander von Humboldt Foundation for
support during his stay in Germany.

\end{document}